# Terahertz driven extremely nonlinear bulk photogalvanic currents in non-resonant conditions

Ofer Neufeld[1,*], Nicolas Tancogne-Dejean[1], Umberto De Giovannini[1,2], Hannes Hübener[1], Angel Rubio[1,3]

[1]Max Planck Institute for the Structure and Dynamics of Matter and Center for Free-Electron Laser Science, Hamburg, Germany, 22761.
[2]IKERBASQUE, Basque Foundation for Science, E-48011, Bilbao, Spain.
[3]Center for Computational Quantum Physics (CCQ), The Flatiron Institute, New York, NY, USA, 10010.

*Corresponding authors E-mails: ofer.neufeld@gmail.com, angel.rubio@mpsd.mpg.de

We report on the generation of bulk photocurrents in materials driven by non-resonant bi-chromatic fields that are circularly polarized and co-rotating. The non-linear photocurrents have a fully controllable directionality and amplitude without requiring carrier-envelope-phase stabilization or few-cycle pulses, and are generated with photon energies much smaller than the band gap (reducing heating in the photo-conversion process). We demonstrate with *ab-initio* calculations that the photocurrent generation mechanism is universal and arises in gaped materials (Si, diamond, MgO, hBN), in semi-metals (graphene), and in two- and three-dimensional systems. Photocurrents are shown to rely on sub-laser-cycle asymmetries in the nonlinear response that build-up coherently from cycle-to-cycle as the conduction band is populated. Importantly, the photocurrents are always transverse to the major axis of the co-circular lasers regardless of the material's structure and orientation (analogously to a Hall current), which we find originates from a generalized time-reversal symmetry in the driven system. At high laser powers ($\sim 10^{13}$ W/cm$^2$) this symmetry can be spontaneously broken by vast electronic excitations, which is accompanied by an onset of carrier-envelope-phase sensitivity and ultrafast many-body effects. Our results are directly applicable for efficient light-driven control of electronics, and for enhancing sub-band-gap bulk photovoltaic effects.

Light-driven dynamics in solids with femtosecond and sub-femtosecond resolution have recently attracted considerable attention. Light-matter interactions can result in novel effects that originate from ultrafast dynamics including high harmonic generation (HHG) [1], and the creation of new states of matter [2–10]. The ability to control the electron motion in real-space, momentum-space and time, can give rise to unprecedented control over observable properties such as light emission [11–19] and magnetic fields [20]. One main avenue of research here is the generation and characterization of light-driven bulk electric currents in the absence of external bias. In materials with broken inversion symmetry, second-order nonlinear effects lead to shift currents through the bulk photovoltaic effect [21,22]. The driving force for carrier separation in the shift current mechanism is the coherent evolution of electron and hole wavefunctions, such that above-bandgap photovoltages can surpass the Shockley–Queisser limit [23]. Photocurrents can also arise in inversion-symmetric materials (where they are standardly forbidden) *via* mixing of bi-chromatic carrier waves with frequencies ω and 2ω [20,24–28]. Here the photon energies are resonant with 2$^{nd}$- and 3$^{rd}$-order perturbative transitions that interfere, and the inversion symmetry is effectively broken by the two-color field (making the effect highly sensitive to the two-color relative phase). The resonant and perturbative nature of these effects precludes access to ultrafast dynamics and possible applications in the Terahertz regime, and is also limited in its conversion efficiency [29].

More recently, nonlinear photocurrents were predicted and observed in dielectrics [30–32] and graphene [33–36] driven by intense quasi-monochromatic few-cycle pulses. The mechanism creating these photocurrents relies on the vector potential of the light field to have a nonzero time integral [32], i.e. there are residual direct terms that only exist in few-cycle pulses. Consequently, these photocurrents are highly sensitive to the carrier-envelope-phase (CEP) and cancel out in non-CEP-stabilized conditions. Controlling or utilizing such currents for ultrafast spectroscopy is challenging, and their applications for energy conversion remains elusive.

Here we explore strong-field driven photocurrents with non-resonant ω-2ω lasers that are circularly-polarized and co-rotating (see Fig. 1). We show that in this regime bulk photocurrents are produced with several attractive features: (i) they do not require CEP stabilized few-cycle pulses, (ii) they do not require resonant or near-resonant transitions, (iii) their amplitude is insensitive to the two-color phase and is finely-tunable by changing the pulse duration, (iv) their directionality is controlled by the two-color phase regardless of the material system. The photocurrents are shown to arise from excited-state occupation imbalances in the Brillouin zone (BZ) that build-up from cycle-to-cycle, and analogously to anomalous Hall currents [9] are always transverse to the main axis of the two-color field. However, in contrast to anomalous Hall currents no bias or magnetic field is required in our case. We explain the origin of this effect from intuitive physical arguments based on electron sub-cycle dynamics, and by a generalized time-reversal symmetry in the laser-dressed system. Lastly, we show that for higher laser powers the generalized time-reversal symmetry is spontaneously broken by strong excitations to the conduction band (CB) that cause a breakdown of the independent-particle approximation, and also lead to CEP sensitivity. These results should be useful for all-optical ultrafast control of electron dynamics, for energy conversion with sub-band-gap light in bulk and spatially uniform materials, and for developing novel ultrafast spectroscopy techniques based on nonlinear photocurrents.

We begin by introducing the ω-2ω laser field that comprises two co-rotating circularly polarized beams:

$$\mathbf{E}(t) = \frac{E_0}{\sqrt{2}} f(t) \, \text{Re}\{(e^{i\omega t} + \Delta e^{2i\omega t + \phi})\hat{\mathbf{e}}_+\} \qquad (1)$$

where $E_0$ is the electric field amplitude of the ω beam taken here in the range of 0.085-0.85 V/Å (equivalent to laser powers of ~$10^{11}$-$10^{13}$ W/cm$^2$), $f(t)$ is a dimensionless envelope function (see SI for details), ω is the fundamental frequency, $\phi$ is the ω-2ω relative phase, $\Delta$ is the amplitude ratio between the beams, and $\hat{\mathbf{e}}_+$ is a right circularly-polarized unit vector. Note that we have employed the dipole approximation, which is valid in our wavelength ranges. A representative Lissajous curve for the time-dependent polarization of this ω-2ω field is presented in Fig. 1(a). Its main characteristic features are: (i) it exhibits a mirror symmetry along its *x*-axis, (ii) the field peak power and time-dependent polarization substantially differ along the positive and negative ends of the *x*-axis, i.e. at the start and middle of an optical cycle where the two fields constructively/destructively interfere (corresponding to *t=0* and *t=T/2*, respectively, see Fig. 1(a)). Importantly, the direction of the field's mirror axis is CEP-independent (it only depends on $\phi$). The interaction of **E**(t) in eq. (1) with matter is described from first principles by using time-dependent density functional theory (TDDFT) within the adiabatic local density approximation (LDA) in the real-space code, octopus [37–39], with which the time-dependent electronic current expectation value, **J**(t), is calculated (all technical details are delegated to the SI).

As a starting point, we analyze driven electron dynamics in diamond subject to **E**(t) in eq. (1). Diamond represents a benchmark system since it has a simple lattice structure with two atoms per unit cell, and is a wide band gap insulator with an indirect gap of 5.48 eV and a direct gap at the Γ point of 7.3 eV [40,41]. These points easily allow extensive calculations with wavelengths well below the band gap (within the LDA the direct gap is 5.65 eV, slightly lower than the experimental value). Figure 1(b) presents the calculated current where the with **E**(t) is polarized in the diamond (111) planes and the field is comprised of 2500nm-1250nm light in the strong-field regime ($I_0$=7.5×$10^{12}$ W/cm$^2$). A residual photocurrent is observed along the *y*-axis after the laser pulse ends (axes are denoted with respect to the field's coordinate system). This effect cannot result from CEP-sensitivity since the pulse is of long duration with a zero time integral over the vector potential. Moreover, the driving frequencies are much below the band gap – it would take over 11 ω photons to induce a transition from the valence band (VB) to the CB. Thus, this effect substantially differs from previously discussed perturbative and/or resonant phenomena [24–28]. Importantly, the current does not decay in our simulation (see Fig. 1(b)). This is true even in the presence of electron-electron interactions incorporated in the simulation, indicating that electrons occupy a true excited state of the system (similar behavior occurs with a Perdew-Burke-Enzerhof (PBE) functional [42], see SI). The current is thus only expected to decay due to scattering channels not included in our model [43].

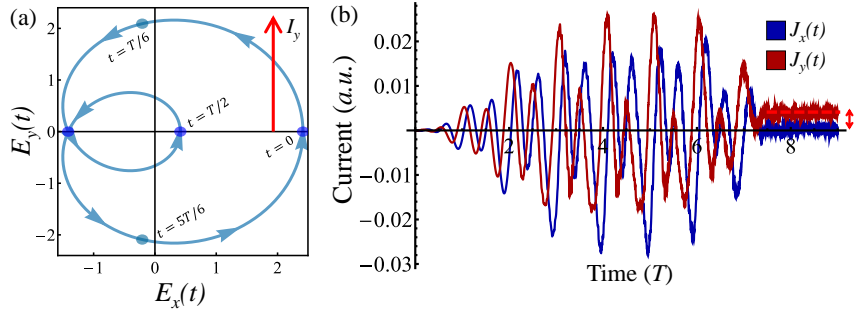

**Figure 1.** Bulk photocurrent generation in diamond by THz co-circular ω-2ω fields. (a) Schematic Lissajous curve for the time-dependent polarization of the field in eq. (1) (for $\phi=0$, $\Delta=\sqrt{2}$). Small arrows indicate the direction of time, and the red arrow indicates the direction of the photocurrent (*I*). Different instances in time are highlighted to emphasize the mirror symmetry of the Lissajous about its *x*-axis, but not the *y*-axis. (b) Calculated current expectation value (see SI for details) for field polarization is in the (111) planes (with $\lambda=2500$nm, $\phi=0$, $\Delta=\sqrt{2}$, $I_0=7.5\times10^{12}$ W/cm$^2$).

To understand the mechanism behind this effect we explore its dependence on the various laser parameters. Figure 2(a) presents the calculated photocurrent amplitude *vs.* the applied laser power, $I_0$, which multiplies both the ω and 2ω components. As expected, the current originates from nonlinear interactions – there is an onset at $I_0=10^{12}$ W/cm$^2$ which increases initially parabolically with $I_0$, indicating that the current is proportional to four-photon processes in the perturbative regime. Indeed, four photons are the minimal number to traverse the band gap for this wavelength (1200nm) while still mixing both photon types (corresponding to annihilation of either three 2ω photons and one ω photon, or two 2ω photons and two ω photons). This mixing is crucial since without it there is no symmetry breaking as in Fig. 1(a) that allows photocurrents to be created (because the field components are individually circularly-polarized with no preferential direction in space).

For powers above $I_0\approx5\times10^{12}$ W/cm$^2$ in Fig. 2(a) the parabolic behavior changes to linear proportionality to $I_0$, paradoxically suggesting that two-photon processes are dominant. Since two photons cannot promote electrons to the CB, higher nonlinear effects are at play. Further increasing the power changes this behavior to an unpredictable pattern that also induces nonzero *x*-polarized currents. We will explore this behavior in detail below, but first we will explain the origin of the photocurrent in the intermediate strong-field regime (for $10^{12}<I_0<10^{13}$ W/cm$^2$). Figure 2(b) presents the current direction *vs.* the orientation of the major axis of **E**(t) in the (111) planes. The photocurrent remains transverse to the *x*-axis defined by the Lissajous in Fig. 1(a). This is despite the fact that different local lattice symmetries are probed as the field rotates (e.g. breaking local reflection or rotational symmetries). Figure 2(b) further shows that throughout this control, the current maintains its amplitude and only exhibits mild modulations. Similar results are obtained when **E**(t) is polarized in the (110) planes that have a different local symmetry (4-fold rather than 6-fold, see SI). These features thus seem analogous to those of anomalous Hall currents, though notably here the laser field has components along both *x* and *y* directions – it is just the main mirror axis of the Lissajous that dictates the direction of the photocurrent.

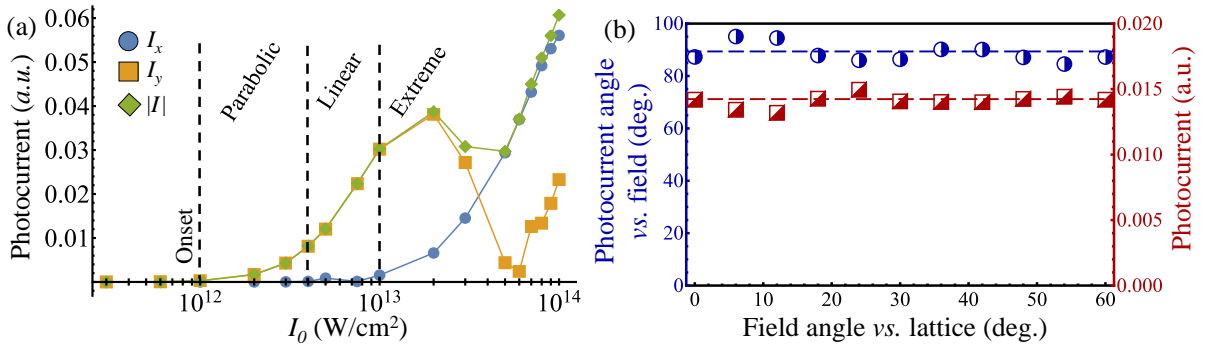

**Figure 2.** Numerical investigation of nonlinear photocurrents in diamond. (a) Photocurrent amplitude *vs.* the field power (for $\lambda=1200$nm, $\phi=0$, $\Delta=\sqrt{2}$). The plot is partitioned to different regimes as indicated by dashed lines. $I_0$ is given in logarithmic scale. (b) Photocurrent direction in the (111) planes w.r.t the field's *x*-axis (blue) and photocurrent amplitude (red) *vs.* the field's orientation w.r.t the lattice (for $\lambda=1200$nm, $\Delta=1$, $I_0=7.5\times10^{12}$ W/cm$^2$).

Figure 3(a,b) presents the integrated and normalized electron occupations in the CB after the laser pulse ends. That is, it shows the occupations along $k_x'$,$k_y'$,$k_z'$ axes in $k$-space after integrating over the other two axes (see SI for details). Note that the 'prime' coordinate system denotes the diamond primitive unit cell coordinates rather than those of the Lissajous, thus a current along the $y$-axis of the field is expected to show up as imbalances in occupations along all three $k_x'$,$k_y'$,$k_z'$ axes. The occupations around the Γ point are asymmetric – electron occupations break the inversion symmetry in $k$-space which is generally a consequence of time-reversal symmetry in the Hamiltonian. Since **E**(t) breaks time-reversal symmetry, $k$-inversion symmetry is broken as expected, allowing photocurrents to be generated. We also note that almost all $k$-points in the BZ contribute to this asymmetry, even far away from the Γ point.

To understand why photocurrent components parallel to the $x$-axis are not generated, we analyze the dynamics of electrons driven in the strong-field regime assuming that the current is generated by two sequential steps: (i) First, the strong laser field excites an electron from the VB to the CB. This step is most likely to occur near the field peak power and near the Γ point (see highlighted points in Fig. 1(a)). (ii) Second, the excited electron is driven in the CB by the laser field, generating intraband currents and occupying different regions in the BZ. Since the Lissajous in Fig. 1(a) is symmetric about its $x$-axis, one expects that roughly an equivalent occupation of states with positive/negative $k_y$ is formed during the acceleration in step (ii). However, because the events that initiate tunneling are very different, a discrepancy is formed in step (i) – trajectories that are initiated for positive $k_y$ are favorable, leading to a residual current. Since no such discrepancies exists along the $y$-axis of the field (e.g., at $t=T/6$ and $t=5T/6$, see Fig. 1(a)), no current is generated along the $x$-axis.

A different explanation for this effect follows from the analysis of the spatio-temporal symmetries of **E**(t) [44]. Due to its mirror axis (see Fig. 1(a)), **E**(t) is invariant under a coupled operation of time-reversal accompanied by a reflection of its $y$-polarization: **E**(t)=σ$_x$×**E**(-t) (here σ$_x$ denotes a reflection $E_y$→-$E_y$). Since for $k$-space occupations time-reversal is equivalent to inversion, this symmetry dictates a mirror plane along the $k_{yz}$-plane (the product of inversion and σ$_x$). That is, occupations must be symmetric for $k_x$→-$k_x$ due to a generalized time-reversal exhibited by the field, forbidding $x$-polarized photocurrents (note that '$x$' labels the coordinate system of the field). We note that this analysis is formally valid only if neglecting other effects that may lead to symmetry breaking, e.g. the finite pulse duration or the local lattice structure. Nevertheless, the numerical results are consistent with this conclusion for a wide parameter range and different materials. This indicates that in this regime the structure of the laser field is crucial.

Figure 3(c) shows the calculated photocurrent amplitudes for different pulse durations for a fixed laser power. The current increases linearly with the pulse duration – the asymmetries in $k$-space occupations build-up coherently from cycle-to-cycle, allowing a finely-tuned control over the current and further emphasizing the importance of sub-cycle dynamics.

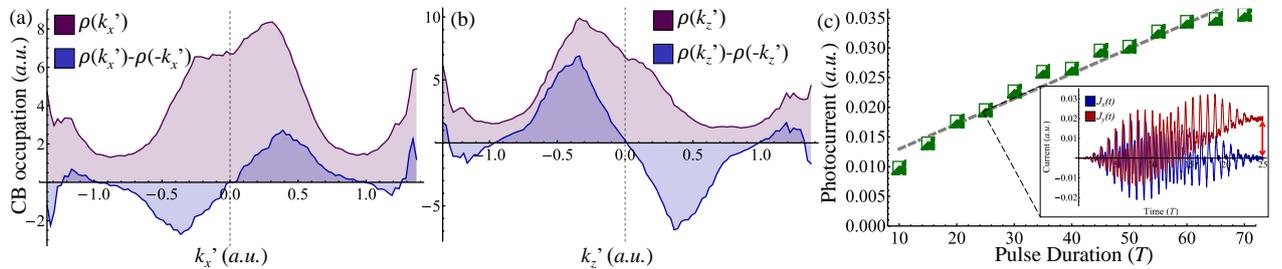

**Figure 3.** Conduction band occupations in diamond after interaction with ω-2ω co-circular laser, and photocurrent proportionality to the pulse duration. (a,b) CB occupations (denoted by ρ) along $k_x'$ and $k_z'$ axes, respectively, integrated over the other two axes, presented in the coordinate system of the lattice ($k_y'$ is not shown due to similarity to $k_x'$). ρ is normalized to the density of states, g($k'$). Purple curves show the CB occupation, while blue curves show the deviation from inversion symmetry (i.e. the occupations subtracted from their inverted image), indicating which states contribute to the photocurrent. Calculated for λ=1200nm, ϕ=0, Δ=√3, $I_0$=7.5×10$^{12}$ W/cm$^2$. (c) Photocurrent amplitude *vs.* pulse duration. Inset shows calculation for a 25-cycle pulse showing the build-up of photocurrent from cycle-to-cycle. Calculated for the same parameters as (a) except Δ=√2.

Having established the main results in diamond, we explore other systems (most results are delegated to the SI for brevity). First, we show that similar effects arise in silicon, a common photovoltaic energy conversion material [45] (see Fig. S4(a) in the SI). Second, we show that nonlinear photocurrents arise also in MgO (see Fig. S4(b) in SI), which is a wide band gap oxide with primarily ionic bonding. Hence, the current generation is largely independent of the solid's characteristics. Third, we investigate a monolayer of hBN subject to **E**(t), showing that similar photocurrents arise (see Fig. S5 in SI). We note that this result differs from the ones obtained with a counter-rotating bi-circular field that does not generate a photocurrent [19,46]. Lastly, we explore driven electron dynamics in graphene. The interactions in graphene are resonant due to its metallic nature. Nonetheless, photocurrents arise just as has been shown experimentally for the few-cycle case [33,34], but with stronger asymmetries in CB occupations (see Fig. S6 in the SI). Remarkably, the K and K' points in the BZ are still connected by a generalized time-reversal symmetry, such that their total integrated occupations are identical, but the occupation patterns allow a photocurrent to be generated.

Finally, we discuss the extreme regime for $I_0 > 10^{13}$ W/cm$^2$ (which is experimentally accessible) in diamond (Fig. 2(a)). Here *x*-polarized photocurrents are also allowed. Moreover, the current increase *vs.* $I_0$ is erratic and indicative of highly nonlinear effects. We find that several phenomena occur that contribute to this behavior and result from spontaneous symmetry breaking to the generalized time-reversal symmetry (that forbids the *x*-polarized currents for $I_0 < 10^{13}$ W/cm$^2$). In this regime the excitation from the first VB to the CB is so fast and vast that the VB maxima at the Γ point has largely depleted even before the laser pulse ends (see Fig. 4(a)). Thus, as the system evolves electrons must traverse larger gaps that are contributed from farther *k*-points, which breaks time-reversal symmetry. Intuitively, this leads to a sub-cycle asymmetry that allows *x*-polarized photocurrents, since occupations in the first half of an optical cycle no-longer exactly cancel with those in the second half. In an extreme case, the VB may be fully depleted in a single half-cycle, resulting in very large *x*-photocurrents. We note that it is unclear if all, or just some, of these effects are observable below the material damage threshold (which can be manipulated by changing the laser parameters or the material).

Besides the onset of *x*-polarized photocurrents, two other effects come into play. First, dynamical electron-electron interactions play a major role. The excitation of many electrons to the CB significantly renormalizes the effective exchange and correlation interactions from cycle-to-cycle. Figure 4(b) shows photocurrent amplitudes calculated by two different levels of theory – either freezing or not the Hartree and exchange-correlation (HXC) potentials to their ground-state form (the frozen HXC approximation is equivalent to the independent-particle approximation (IPA)). Discrepancies between these theories only arise due to the inclusion of electron-electron interactions induced by the laser, and are usually very small [47–52]. Below $10^{13}$ W/cm$^2$ these are negligible (<5%). Above $10^{13}$ W/cm$^2$ there are significant deviations in the obtained photocurrent (>20%, see Fig. 4(b)), as well as in the intensity of harmonic emission (>50%, see Fig. S3 in the SI). The deviations increase with the field strength, in accordance with the CB population (e.g. compare Fig. 4(a) to Fig. 4(b)). We note that it is not yet clear whether or not this effect is well-described within adiabatic TDDFT, thus results could be affected by the level of theory describing electron-electron interactions. Second, otherwise negligible envelope effects become significant. Figure 4(c) shows the photocurrent amplitudes in this regime *vs.* the CEP, validating that there is strong CEP sensitivity even though the pulse durations are relatively long, and no such sensitivity is observed for weaker power (see SI Fig. S2). Both of these phenomena could be probed by measuring the photocurrent and its direction in space.

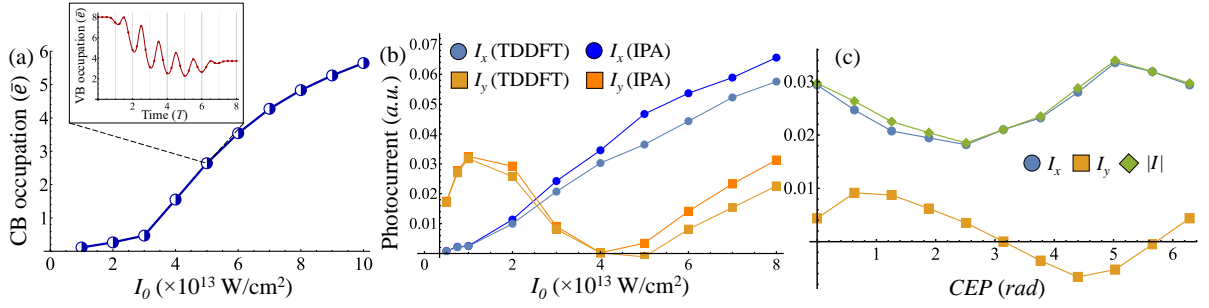

**Figure 4.** Spontaneous symmetry breaking in extremely nonlinear photocurrents in diamond. (a) Total excitation of electrons to the CB per primitive cell *vs.* $I_0$. Inset shows the electron occupation on the VBs during interaction with the laser pulse for the exemplary case of $I_0=5\times10^{13}$ W/cm$^2$, showing that the VB is quickly depopulated even during the turn-on of the laser pulse. (b) Deviation in photocurrent amplitudes between the IPA and the TDDFT calculation *vs.* field power (for $\lambda=1200$nm, $\phi=0$, $\Delta=\sqrt{3}$, $I_0=5\times10^{13}$ W/cm$^2$). (c) Calculated photocurrent amplitudes *vs.* CEP, showing that CEP-sensitivity arises in the extremely strong-field regime. Calculated for the same parameters as (a), except with $\Delta=\sqrt{2}$.

To summarize, we have investigated the interaction of solids with a bi-chromatic ω-2ω laser with co-rotating circularly-polarized components. We showed that photocurrents are generated in the bulk without external bias, and due to a coherent excitation of electrons to the CB that builds-up from cycle-to-cycle. A generalized time-reversal symmetry exhibited in the system imposes that only transverse photocurrents are generated, allowing ideal control over the current amplitude and directionality. This effect is general to three- and two-dimensional systems, gapped insulators, semi-metals, and does not strongly depend on the material properties. We also analyzed an extreme strong-field regime where time-reversal is spontaneously broken, leading to ultrafast laser-induced correlations, as well as CEP-sensitivity from long pulses. Both these phenomena pave the way to novel forms of ultrafast spectroscopy for electron dynamics that rely on nonlinear photocurrents. Importantly, our results demonstrate a method for sub-band-gap bulk photovoltaic energy conversion that is not bounded by any theoretical limit. Looking forward, these ideas could also be employed to enhance the photogalvanic effect in perovskites [22], and for improved control in petahertz electronics [53].

## ASSOCIATED CONTENT

Supplemental material is available for this paper that includes details on the methodology used in calculations as well as some additional results.

## ACKNOWLEDGMENTS

We thank Shunsuke A. Sato from University of Tsukuba for helpful discussions. We acknowledge financial support from the European Research Council (ERC-2015-AdG-694097). The Flatiron Institute is a division of the Simons Foundation. O.N. gratefully acknowledges the support of the Humboldt foundation. This work was supported by the Cluster of Excellence Advanced Imaging of Matter (AIM), Grupos Consolidados (IT1249-19) and SFB925.

## SUPPLEMENTARY INFORMATION

## I. COMPUTATIONAL DETAILS

We report here on technical details for the calculations presented in the main text. We start by presenting the methodological approach utilized in the main text that is based on time-dependent density functional theory (TDDFT). All DFT calculations were performed using the real-space grid-based code, octopus [37–39]. The Kohn Sham (KS) equations were discretized on a Cartesian grid with the shape of the primitive lattice cell in each material system, where atomic geometries and lattice parameters were taken at the experimental values. Calculations were performed using the local density approximation (LDA), and in some cases also using the Perdew–Burke–Ernzerhof (PBE) exchange-correlation (XC) functional [42]. Spin degrees of freedom and spin-orbit couplings were neglected. The frozen core approximation was used for inner core bands which were treated with norm-conserving pseudopotentials [54]. The KS equations were solved to self-consistency with a tolerance $<10^{-7}$ Hartree, and the grid spacing was converged in each material system to Δ*x*=Δ*y*=Δ*z*=0.33 Bohr in diamond, 0.35 Bohr in Si, 0.3 bohr in MgO, 0.4 Bohr in monolayer hBN, and 0.4 Bohr in graphene. We employed a Γ-centered *k*-grid in each system, which converged the time-dependent current expectation value: 36×36×36 *k*-grid in diamond, 24×24×24 *k*-grid in Si, 32×32×32 *k*-grid in MgO, 36×36×1 *k*-grid in monolayer hBN, and 120×120×1 *k*-grid in graphene.

For TDDFT calculations, we solved the time-dependent KS equations within the adiabatic approximation, represented in real-space and in the velocity gauge, given in atomic units by:

$$i\partial_t |\varphi_{n,k}^{KS}(t)\rangle = \left(\frac{1}{2}\left(-i\nabla + \frac{\mathbf{A}(t)}{c}\right)^2 + v_{KS}(\mathbf{r},t)\right) |\varphi_{n,k}^{KS}(t)\rangle \tag{S1}$$

where $|\varphi_{n,k}^{KS}(t)\rangle$ is the KS-Bloch state at *k*-point *k* and band *n*, $\mathbf{A}(t)$ is the vector potential of the laser electric field within the dipole approximation, such that $-\partial_t \mathbf{A}(t) = c\mathbf{E}(t)$, *c* is the speed of light in atomic units (*c*≈137.036), and $v_{KS}(\mathbf{r},t)$ is the time-dependent KS potential given by:

$$v_{KS}(\mathbf{r},t) = -\sum_I \frac{Z_I}{|\mathbf{R}_I - \mathbf{r}|} + \int d^3 r' \frac{n(\mathbf{r}',t)}{|\mathbf{r}-\mathbf{r}'|} + v_{XC}[n(\mathbf{r},t)] \tag{S2}$$

where $Z_I$ is the charge of the *I*'th nuclei and $\mathbf{R}_I$ is its coordinate, $v_{XC}$ is the XC potential that is a functional of $n(\mathbf{r},t)=\sum_{n,k}\left|\varphi_{n,k}^{KS}(t)\right|^2$, the time-dependent electron density. The KS wave functions were propagated with a time step of Δ*t*=0.2 a.u. which converged the time-dependent current in all material systems within the LDA. For TD-PBE calculations a time step of 0.08 a.u. was used. The initial state was taken to be the system's ground state. The propagator was represented by a Lanczos expansion.

For the independent particle approximation (IPA) calculations, the same methodology was utilized but where the KS potential was kept frozen to its initial form such that $v_{KS}[n(\mathbf{r},t)] \equiv v_{KS}[n(\mathbf{r},t=0)]$.

For the two-dimensional systems of monolayer hBN and graphene, we added additional vacuum spacing above and below the monolayer of 40 Bohr in each direction of the non-periodic axis, and absorbing boundaries were employed along this axis with a width of 12 Bohr.

The time-dependent current expectation value was calculated directly from the time-dependent KS states as:

$$\mathbf{J}(t) = \int_\Omega d^3 r \, \mathbf{j}(\mathbf{r},t) \tag{S3}$$

where $\mathbf{j}(\mathbf{r},t)$ is the microscopic time-dependent current density:

$$\mathbf{j}(\mathbf{r},t) = \frac{1}{2}\sum_{n,k}\left[\varphi_{n,k}^{KS\,*}(r,t)\left(-i\nabla + \frac{\mathbf{A}(t)}{c}\right)\varphi_{n,k}^{KS}(r,t) + c.c.\right] \tag{S4}$$

, and Ω represents the volume integral over the primitive cell. The photocurrent was calculated as $\mathbf{I} = \mathbf{J}(t \to \infty)$ when the laser pulse has ended. For results presented in the SI, the harmonic spectrum was calculated as the Fourier transform of the first derivative of the current:

$$I(\omega) = \left|\int dt\, \partial_t \mathbf{J}(t) e^{-i\omega t}\right|^2 \tag{S5}$$

, which was evaluated numerically with an 8'th order finite-difference approximation for the temporal derivative, and fast Fourier transforms (FFT).

Band and *k*-point occupations were computed by calculating the projections of the time-dependent KS states on the field-free states at *t*=0. For Figures 3(a,b) in the main text, the occupations were integrated in *k*-space and normalized to the density of states at each *k*-point, g(*k*).

The envelope function of the employed laser pulse, *f(t)*, was taken to be of the following 'super-sine' form [51]:

$$f(t) = \left(sin\left(\pi\frac{t}{T_p}\right)\right)^{\left(\frac{\left|\pi\left(\frac{t}{T_p}-\frac{1}{2}\right)\right|}{\sigma}\right)} \tag{S6}$$

where $\sigma=0.75$, $T_p$ is the duration of the laser pulse which was taken to be $T_p=8T$ (unless stated otherwise), where $T$ is a single cycle of the fundamental carrier frequency. This form is roughly equivalent to a super-gaussian pulse, but where the field starts and ends exactly at 0 which is more convenient numerically.

## II. ADDITIONAL RESULTS IN DIAMOND

We present here additional results complementary to those presented in the main text. First, we show that the photocurrents arise even when utilizing a PBE XC functional, and similarly to the TD-LDA calculations presented in the main text, they do not decay over the timescale of the simulation. Figure S1(a) shows the time-dependent current expectation value calculated in diamond with PBE XC. The current arises only along the *y*-axis transverse to the field's mirror plane and remains constant in time even long after the pulse has ended. This supports the conclusions in the main text that the excited current-carrying state is an eigenstate of the many-body system and will only decay *via* radiative or scattering channels that are not included in our simulation.

Next, Fig. S1(b) shows the calculated time-dependent current expectation value driven in a diamond lattice within the (110) planes (unlike figures in the main text that present currents driven in the (111) planes). The (110) planes have a local four-fold symmetry that is different than that in the (111) planes. Still, the current arises along the field's y-axis, demonstrating similar control over the current direction. This supports the conclusion that the current generation mechanism is largely unaffected by the local lattice structure and symmetry.

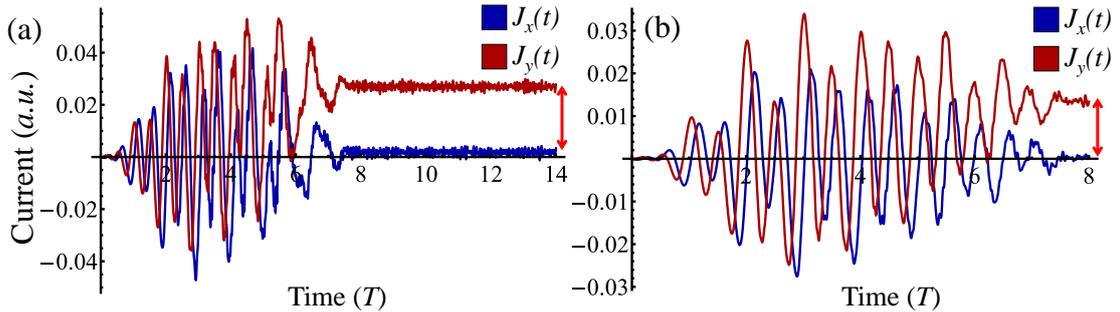

**Figure S1.** Additional calculations of photocurrents in diamond. (a) Time-dependent current expectation value calculated using PBE XC, showing results similar to those obtained within the LDA (for laser parameters $\lambda=1200$nm, $\Delta=\sqrt{3}$, $I_0=7.5\times10^{12}$ W/cm$^2$). (b) Time-dependent current expectation value calculated where **E**(t) is polarized in the diamond (110) planes, showing that that the photocurrent is still directed along the field's *y*-axis regardless of the different local symmetry compared to the (111) planes (calculated for similar laser parameters as in (a), except that $\Delta=1$). Arrows indicate the residual photocurrent.

We next show that these currents are CEP-independent, just as one would expect for long-duration pulses. Fig. S2 shows calculated photocurrent amplitudes in diamond vs. the CEP for the same pulse durations as presented in the main text in Fig. 4(c). The resulting photocurrents are CEP-independent. This result further highlights that different physical effects come into play in higher laser powers, as discussed in the main text.

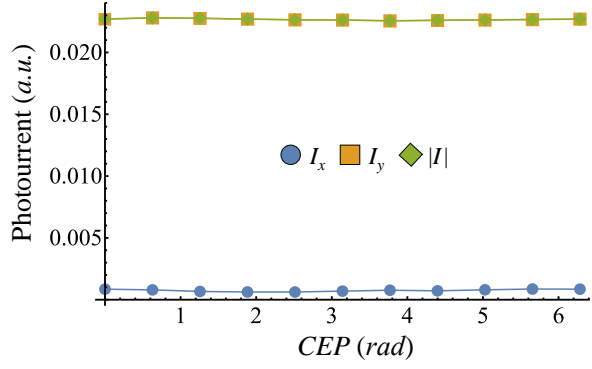

**Figure S2.** CEP-independence of photocurrents in diamond for laser powers smaller than $10^{13}$ W/cm$^2$. Driven photocurrent amplitudes in diamond (111) planes vs. the CEP (calculated for laser parameters λ=1200nm, Δ=√2, $I_0$=7.5×10$^{12}$ W/cm$^2$ and 8-cycle long laser pulses).

Lastly, we explore here deviations from the IPA for strongly-driven diamond in the regime of laser powers above $10^{13}$ W/cm$^2$. Results in the main text (Fig. 4(b)) showed that there are considerable deviations in the obtained photocurrents upon inclusion/exclusion of dynamical electron-electron interactions (up to 20%), as well as an onset of *x*-polarized photocurrents. We show here that in these conditions similar effects arise in the HHG emission. In this case deviations are even larger and can reach values of 100% for some harmonics, indicating the breakdown of the IPA. Figure S3(a) presents the calculated HHG spectra in this regime from the co-circular ω-2ω field for $I_0$=5×10$^{13}$ W/cm$^2$, showing these extreme deviations. Figure S3(b) shows that when averaged over the first 60 harmonic orders, these deviations increase in correspondence with the laser power, similar to the photocurrent deviations. Overall, this further validates the onset of strong laser-induced many-body effects in this intense regime.

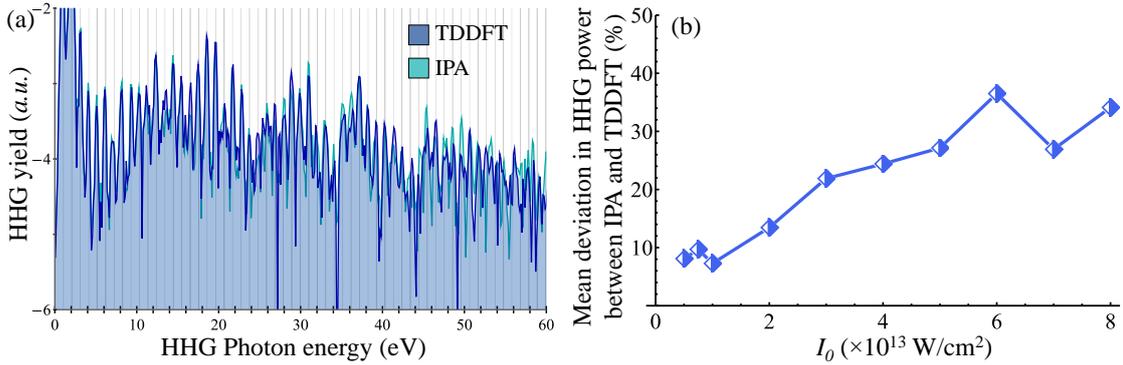

**Figure S3.** Breakdown of the IPA in strongly-driven diamond in the HHG spectrum. (a) Calculated HHG spectra within the IPA and full TDDFT calculation for laser conditions λ=1200nm, Δ=√3, $I_0$=5×10$^{13}$ W/cm$^2$. Gray lines indicate positions of integer harmonic orders for this wavelength. The plot shows that considerable deviations arise in the HHG power in this regime due to dynamical correlations. (b) Average deviation in harmonic power between the full TDDFT calculation and the IPA *vs.* the laser power (averaged over first 60 harmonics).

## III. ADDITIONAL RESULTS IN OTHER SYSTEMS

We present additional results of photocurrent excitation in other material systems. We start by exploring several three-dimensional bulk solids. Figure S4 presents calculated time-dependent photocurrents in Si (in the (111) planes), and MgO (in the (111) planes), after interaction with intense co-circular ω-2ω pulses. These results are analogous to those presented in the main text for diamond and highlight that the photocurrent and its generation mechanism is largely independent of the material system and its chemical and physical properties. In particular, the current is always observed transverse to the field's mirror axis, and is generated even though the laser pulses are of long duration and have photon energies much smaller than the band gap in each case.

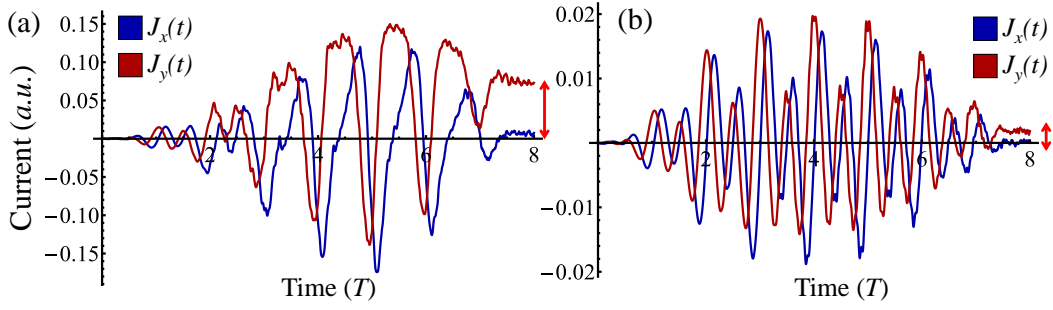

**Figure S4.** Calculations of photocurrents in Si and MgO. (a) Time-dependent current expectation value in Si (for laser parameters λ=2500nm, Δ=1, $I_0=10^{12}$ W/cm$^2$). (b) Time-dependent current expectation value in MgO (for laser parameters λ=1200nm, Δ=1, $I_0=5\times10^{12}$ W/cm$^2$). Arrows indicate the residual photocurrent.

We next investigate photocurrents in two-dimensional systems. Figure S5 presents investigation of photocurrent generation in monolayer hBN where the field is polarized within the monolayer (*xy* plane). From Fig. S5(a) we can tell that similar photocurrents arise, and that they are transverse to the field's mirror axis. We note that this occurs even though hBN monolayers lack inversion symmetry. Figure S5(b) shows the *k*-space occupations of the CB after the laser pulse has ended, clearly indicating that there is no *k*-inversion symmetry (i.e. a residual current is generated). Notably, in this case no symmetries remain in the *k*-space occupations, but the current is still mainly polarized along the *y*-axis, indicating that its directionality is largely determined by the field's properties rather than those of the material system. We also note that, as expected, the occupations of the K and K' valleys are now asymmetric, allowing for a mechanism to generate valley polarization. This is in accordance with similar results presented with counter-rotating bi-circular fields [19].

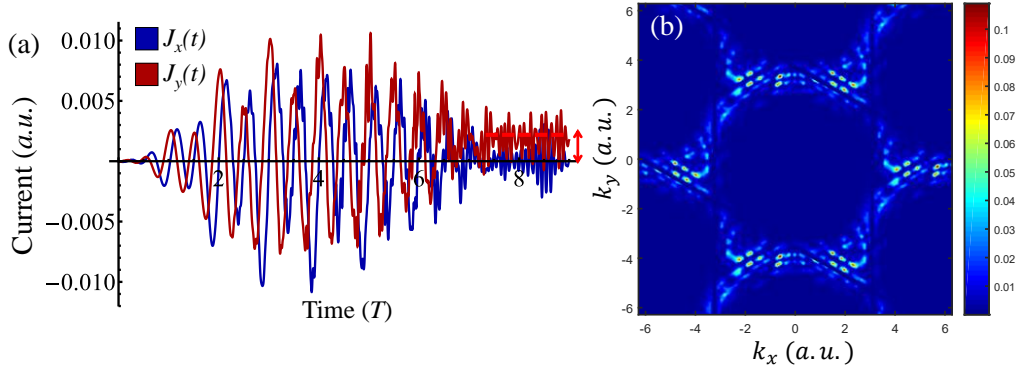

**Figure S5.** Photocurrent generation in monolayer hBN. (a) Time-dependent current expectation value calculated for the laser parameters λ=2500nm, Δ=√2, $I_0=10^{12}$ W/cm$^2$. Arrow and dashed line indicate the residual photocurrent. (b) CB occupations in *k*-space after the pulse has ended showing an asymmetric occupation pattern around the K and K' points.

We move on to graphene, which is a hexagonal Dirac semi-metal with band touching at the K and K' points. Previous theoretical and experimental work showed that one can generate strongly-driven photocurrents in graphene with monochromatic few-cycle laser pulses [33–36]. Figure S6 explores current injection from the co-rotating ω-2ω field, showing that similar effects occur in the long-pulse regime (the field is polarized within the graphene sheet). This effect is seen both for shorter wavelengths (1200-600nm, see Fig. S6(a,b)), and for much longer wavelengths in the THz regime (2500-1250nm, see Fig. S6(c,d)). We note that for all tested wavelengths the *k*-space occupations exhibit an exact mirror symmetry, in accordance with the formal analysis presented in the main text (and because graphene is inversion-symmetric). Thus, the total populations of the K and K' valleys are identical and connected through a mirror symmetry, which is mediated by the generalized time-reversal symmetry of the ω-2ω co-rotating field. Nonetheless, photocurrents arise because the occupation patterns are extremely asymmetric between the K and K' valleys (this is also evident locally in the region near each K and the K' band touching point).

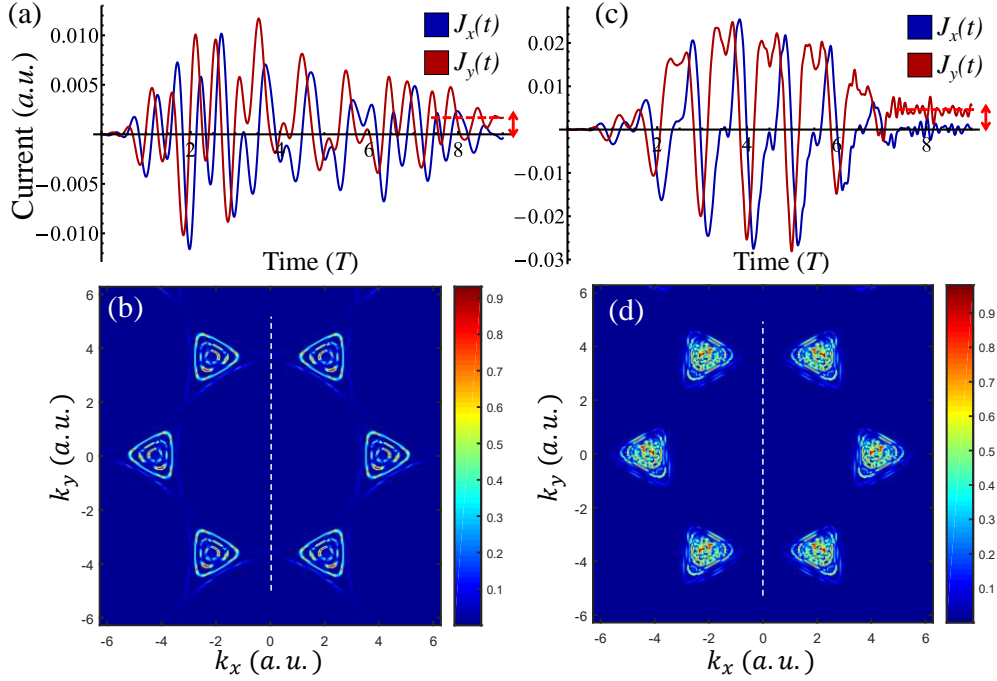

**Figure S6.** Photocurrent generation in graphene. (a) Time-dependent current expectation value calculated for the laser parameters $\lambda=1200$nm, $\Delta=\sqrt{2}$, $I_0=10^{11}$ W/cm$^2$). (b) CB occupations in *k*-space after the pulse has ended corresponding to the parameters in (a). (c,d) Same as in (a,b), but for $\lambda=2500$nm. Arrows and dashed lines indicate the residual photocurrents. White dashed lines mark a mirror symmetry for *k*-space occupations that connects the K and K' points, and forbids *x*-polarized photocurrents.

## REFERENCES

[1] S. Ghimire and D. A. Reis, *High-Harmonic Generation from Solids*, Nat. Phys. **15**, 10 (2019).

[2] N. H. Lindner, G. Refael, and V. Galitski, *Floquet Topological Insulator in Semiconductor Quantum Wells*, Nat. Phys. **7**, 490 (2010).

[3] Y. H. Wang, H. Steinberg, P. Jarillo-Herrero, and N. Gedik, *Observation of Floquet-Bloch States on the Surface of a Topological Insulator*, Science **342**, 453 LP (2013).

[4] A. G. Grushin, Á. Gómez-León, and T. Neupert, *Floquet Fractional Chern Insulators*, Phys. Rev. Lett. **112**, 156801 (2014).

[5] P. Titum, N. H. Lindner, M. C. Rechtsman, and G. Refael, *Disorder-Induced Floquet Topological Insulators*, Phys. Rev. Lett. **114**, 056801 (2015).

[6] H. Hübener, M. A. Sentef, U. De Giovannini, A. F. Kemper, and A. Rubio, *Creating Stable Floquet-Weyl Semimetals by Laser-Driving of 3D Dirac Materials*, Nat. Commun. **8**, 13940 (2017).

[7] H. Hübener, U. De Giovannini, and A. Rubio, *Phonon Driven Floquet Matter*, Nano Lett. **18**, 1535 (2018).

[8] G. E. Topp, G. Jotzu, J. W. McIver, L. Xian, A. Rubio, and M. A. Sentef, *Topological Floquet Engineering of Twisted Bilayer Graphene*, Phys. Rev. Res. **1**, 1 (2019).

[9] J. W. McIver, B. Schulte, F.-U. Stein, T. Matsuyama, G. Jotzu, G. Meier, and A. Cavalleri, *Light-Induced Anomalous Hall Effect in Graphene*, Nat. Phys. **16**, 38 (2020).

[10] M. Buzzi, D. Nicoletti, M. Fechner, N. Tancogne-Dejean, M. A. Sentef, A. Georges, T. Biesner, E. Uykur, M. Dressel, A. Henderson, T. Siegrist, J. A. Schlueter, K. Miyagawa, K. Kanoda, M.-S. Nam, A. Ardavan, J. Coulthard, J. Tindall, F. Schlawin, D. Jaksch, and A. Cavalleri, *Photomolecular High-Temperature Superconductivity*, Phys. Rev. X **10**, 031028 (2020).

[11] M. Schultze, E. M. Bothschafter, A. Sommer, S. Holzner, W. Schweinberger, M. Fiess, M. Hofstetter, R. Kienberger, V. Apalkov, V. S. Yakovlev, M. I. Stockman, and F. Krausz, *Controlling Dielectrics with the Electric Field of Light*, Nature **493**, 75 (2013).

[12] O. Schubert, M. Hohenleutner, F. Langer, B. Urbanek, C. Lange, U. Huttner, D. Golde, T. Meier, M. Kira, S. W. Koch, and R. Huber, *Sub-Cycle Control of Terahertz High-Harmonic Generation by Dynamical Bloch Oscillations*, Nat. Photonics **8**, 119 (2014).

[13] H. Liu, Y. Li, Y. S. You, S. Ghimire, T. F. Heinz, and D. A. Reis, *High-Harmonic Generation from an Atomically Thin Semiconductor*, Nat. Phys. **13**, 262 (2017).

[14] Y. S. You, D. A. Reis, and S. Ghimire, *Anisotropic High-Harmonic Generation in Bulk Crystals*, Nat. Phys. **13**, 345 (2017).

[15] T. T. Luu and H. J. Wörner, *Measurement of the Berry Curvature of Solids Using High-Harmonic Spectroscopy*, Nat. Commun. **9**, 916 (2018).

[16] D. Bauer and K. K. Hansen, *High-Harmonic Generation in Solids with and without Topological Edge States*, Phys. Rev. Lett. **120**, 177401 (2018).

[17] K. K. Hansen, D. Bauer, and L. B. Madsen, *Finite-System Effects on High-Order Harmonic Generation: From Atoms to Solids*, Phys. Rev. A **97**, 43424 (2018).

[18] R. E. F. Silva, Á. Jiménez-Galán, B. Amorim, O. Smirnova, and M. Ivanov, *Topological Strong-Field Physics on Sub-Laser-Cycle Timescale*, Nat. Photonics **13**, 849 (2019).

[19] Á. Jiménez-Galán, R. E. F. Silva, O. Smirnova, and M. Ivanov, *Lightwave Control of Topological Properties in 2D Materials for Sub-Cycle and Non-Resonant Valley Manipulation*, Nat. Photonics **14**, 728 (2020).

[20] S. Sederberg, F. Kong, F. Hufnagel, C. Zhang, E. Karimi, and P. B. Corkum, *Vectorized Optoelectronic Control and Metrology in a Semiconductor*, Nat. Photonics **14**, 680 (2020).

[21] I. Grinberg, D. V. West, M. Torres, G. Gou, D. M. Stein, L. Wu, G. Chen, E. M. Gallo, A. R. Akbashev, P. K. Davies, J. E. Spanier, and A. M. Rappe, *Perovskite Oxides for Visible-Light-Absorbing Ferroelectric and Photovoltaic Materials*, Nature **503**, 509 (2013).

[22] L. Z. Tan, F. Zheng, S. M. Young, F. Wang, S. Liu, and A. M. Rappe, *Shift Current Bulk Photovoltaic Effect in Polar Materials—Hybrid and Oxide Perovskites and Beyond*, Npj Comput. Mater. **2**, 16026 (2016).

[23] W. Shockley and H. J. Queisser, *Detailed Balance Limit of Efficiency of P-n Junction Solar Cells*, J. Appl. Phys. **32**, 510 (1961).

[24] R. Atanasov, A. Haché, J. L. P. Hughes, H. M. van Driel, and J. E. Sipe, *Coherent Control of Photocurrent Generation in Bulk Semiconductors*, Phys. Rev. Lett. **76**, 1703 (1996).

[25] A. Haché, Y. Kostoulas, R. Atanasov, J. L. P. Hughes, J. E. Sipe, and H. M. van Driel, *Observation of Coherently Controlled Photocurrent in Unbiased, Bulk GaAs*, Phys. Rev. Lett. **78**, 306 (1997).

[26] R. D. R. Bhat and J. E. Sipe, *Optically Injected Spin Currents in Semiconductors*, Phys. Rev. Lett. **85**, 5432 (2000).

[27] E. Sternemann, T. Jostmeier, C. Ruppert, H. T. Duc, T. Meier, and M. Betz, *Femtosecond Quantum Interference Control of Electrical Currents in GaAs: Signatures beyond the Perturbative Chi(3) Limit*, Phys. Rev. B **88**, 165204 (2013).

[28] D. A. Bas, K. Vargas-Velez, S. Babakiray, T. A. Johnson, P. Borisov, T. D. Stanescu, D. Lederman, and A. D. Bristow, *Coherent Control of Injection Currents in High-Quality Films of Bi2Se3*, Appl. Phys. Lett. **106**, 41109 (2015).

[29] L. Z. Tan and A. M. Rappe, *Upper Limit on Shift Current Generation in Extended Systems*, Phys. Rev. B **100**, 85102 (2019).

[30] A. Schiffrin, T. Paasch-Colberg, N. Karpowicz, V. Apalkov, D. Gerster, S. Mühlbrandt, M. Korbman, J. Reichert, M. Schultze, S. Holzner, J. V Barth, R. Kienberger, R. Ernstorfer, V. S. Yakovlev, M. I. Stockman, and F. Krausz, *Optical-Field-Induced Current in Dielectrics*, Nature **493**, 70 (2013).

[31] M. S. Wismer, S. Y. Kruchinin, M. Ciappina, M. I. Stockman, and V. S. Yakovlev, *Strong-Field Resonant Dynamics in Semiconductors*, Phys. Rev. Lett. **116**, 197401 (2016).

[32] F. Langer, Y.-P. Liu, Z. Ren, V. Flodgren, C. Guo, J. Vogelsang, S. Mikaelsson, I. Sytcevich, J. Ahrens, A. L'Huillier, C. L. Arnold, and A. Mikkelsen, *Few-Cycle Lightwave-Driven Currents in a Semiconductor at High Repetition Rate*, Optica **7**, 276 (2020).

[33] T. Higuchi, C. Heide, K. Ullmann, H. B. Weber, and P. Hommelhoff, *Light-Field-Driven Currents in Graphene*, Nature **550**, 224 (2017).

[34] C. Heide, T. Higuchi, H. B. Weber, and P. Hommelhoff, *Coherent Electron Trajectory Control in Graphene*, Phys. Rev. Lett. **121**, 207401 (2018).

[35] C. Heide, T. Boolakee, T. Higuchi, H. B. Weber, and P. Hommelhoff, *Interaction of Carrier Envelope Phase-Stable Laser Pulses with Graphene: The Transition from the Weak-Field to the Strong-Field Regime*, New J. Phys. **21**, 45003 (2019).

[36] C. Heide, T. Boolakee, T. Higuchi, and P. Hommelhoff, *Sub-Cycle Temporal Evolution of Light-Induced Electron Dynamics in Hexagonal 2D Materials*, J. Phys. Photonics **2**, 24004 (2020).

[37] A. Castro, H. Appel, M. Oliveira, C. A. Rozzi, X. Andrade, F. Lorenzen, M. A. L. Marques, E. K. U. Gross, and A. Rubio, *Octopus: A Tool for the Application of Time-Dependent Density Functional Theory*, Phys. Status Solidi **243**, 2465 (2006).

[38] X. Andrade, D. Strubbe, U. De Giovannini, H. Larsen, M. J. T. Oliveira, J. Alberdi-rodriguez, A. Varas, I. Theophilou, N. Helbig, M. J. Verstraete, L. Stella, F. Nogueira, A. Castro, M. A. L. Marques, and A. Rubio, *Real-Space Grids and the Octopus Code as Tools for the Development of New Simulation Approaches for Electronic Systems*, Phys. Chem. Chem. Phys. **17**, 31371 (2015).

[39] N. Tancogne-Dejean, M. J. T. Oliveira, X. Andrade, H. Appel, C. H. Borca, G. Le Breton, F. Buchholz, A. Castro, S. Corni, A. A. Correa, U. De Giovannini, A. Delgado, F. G. Eich, J. Flick, G. Gil, A. Gomez, N. Helbig, H. Hübener, R. Jestädt, J. Jornet-Somoza, A. H. Larsen, I. V Lebedeva, M. Lüders, M. A. L. Marques, S. T. Ohlmann, S. Pipolo, M. Rampp, C. A. Rozzi, D. A. Strubbe, S. A. Sato, C. Schäfer, I. Theophilou, A. Welden, and A. Rubio, *Octopus, a Computational Framework for Exploring Light-Driven Phenomena and Quantum Dynamics in Extended and Finite Systems*, J. Chem. Phys. **152**, 124119 (2020).

[40] R. A. Roberts and W. C. Walker, *Optical Study of the Electronic Structure of Diamond*, Phys. Rev. **161**, 730 (1967).

[41] H. Löfås, A. Grigoriev, J. Isberg, and R. Ahuja, *Effective Masses and Electronic Structure of Diamond Including Electron Correlation Effects in First Principles Calculations Using the GW-Approximation*, AIP Adv. **1**, 32139 (2011).

[42] J. P. Perdew, K. Burke, and M. Ernzerhof, *Generalized Gradient Approximation Made Simple*, Phys. Rev. Lett. **77**, 3865 (1996).

[43] M. Bernardi, D. Vigil-Fowler, J. Lischner, J. B. Neaton, and S. G. Louie, *Ab Initio Study of Hot Carriers in the First Picosecond after Sunlight Absorption in Silicon*, Phys. Rev. Lett. **112**, 257402 (2014).

[44] O. Neufeld, D. Podolsky, and O. Cohen, *Floquet Group Theory and Its Application to Selection Rules in Harmonic Generation*, Nat. Commun. **10**, 405 (2019).

[45] R. M. Swanson, *A Vision for Crystalline Silicon Photovoltaics*, Prog. Photovoltaics Res. Appl. **14**, 443 (2006).

[46] M. S. Mrudul, Á. Jiménez-Galán, M. Ivanov, and G. Dixit, *Light-Induced Valleytronics in Pristine Graphene*, Optica **8**, 422 (2021).

[47] N. Tancogne-Dejean, O. D. Mücke, F. X. Kärtner, and A. Rubio, *Ellipticity Dependence of High-Harmonic Generation in Solids Originating from Coupled Intraband and Interband Dynamics*, Nat. Commun. **8**, 745 (2017).

[48] I. Floss, C. Lemell, G. Wachter, V. Smejkal, S. A. Sato, X. M. Tong, K. Yabana, and J. Burgdörfer, *Ab Initio Multiscale Simulation of High-Order Harmonic Generation in Solids*, Phys. Rev. A **97**, 011401(R) (2018).

[49] C. Yu, K. K. Hansen, and L. B. Madsen, *High-Order Harmonic Generation in Imperfect Crystals*, Phys. Rev. A **99**, 063408 (2019).

[50] X.-M. Tong and S.-I. Chu, *Time-Dependent Density-Functional Theory for Strong-Field Multiphoton Processes: Application to the Study of the Role of Dynamical Electron Correlation in Multiple High-Order Harmonic Generation*, Phys. Rev. A **57**, 452 (1998).

[51] O. Neufeld and O. Cohen, *Background-Free Measurement of Ring Currents by Symmetry-Breaking High-Harmonic Spectroscopy*, Phys. Rev. Lett. **123**, 103202 (2019).

[52] O. Neufeld and O. Cohen, *Probing Ultrafast Electron Correlations in High Harmonic Generation*, Phys. Rev. Res. **2**, 033037 (2020).

[53] J. Schoetz, Z. Wang, E. Pisanty, M. Lewenstein, M. F. Kling, and M. F. Ciappina, *Perspective on Petahertz Electronics and Attosecond Nanoscopy*, ACS Photonics **6**, 3057 (2019).

[54] C. Hartwigsen, S. Goedecker, and J. Hutter, *Relativistic Separable Dual-Space Gaussian Pseudopotentials from H to Rn*, Phys. Rev. B **58**, 3641 (1998).